\documentclass[iop]{emulateapj}
\begin{document}

\title{Predictions for reverberating spectral line from a newly formed black hole accretion disk: case of tidal disruption flares}
\author{Wenda Zhang, Wenfei Yu} 
\affil{Shanghai Astronomical Observatory and Key Laboratory for
Research in Galaxies and Cosmology,\\ Chinese Academy of Sciences,
80 Nandan Road, Shanghai, 200030, China. E-mail: wenfei@shao.ac.cn}
\author{Vladim\'{\i}r Karas, Michal Dov\v{c}iak}
\affil{Astronomical Institute, Academy of Sciences of the Czech
Republic,\\ Bo\v{c}n\'{\i}~II~1401, CZ-14100~Prague, Czech
Republic. E-mail: vladimir.karas@cuni.cz}
\shorttitle{Reverberation light curves and spectral lines from a black hole accretion disk}
\shortauthors{Wenda Zhang et al.}

\begin{abstract} 
Tidal Disruption Events (TDEs) can be perfect probes of dormant
SMBHs in normal galaxies. During the rising phase, the accretion luminosity can increase by orders of magnitude in several weeks and the emergent ionizing radiation illuminates the fresh accretion flow. In this paper, we simulated the evolution of the expected spectral line profile of iron due to such a flare by using a ray-tracing code with effects of general relativity (GR) taken into account. We found that the time-dependent profile changes significantly with black hole spin, inclination angle with respect to the black-hole equatorial plane, and the expansion velocity of the ionization front. At low values of spin, a ``loop'' feature appears in the line profile vs.\ time plot when the inclination is no less than $30^\circ$ and the expansion velocity $v_{\rm exp}$ is no less than half speed of light, due to a shadow in the emission of the truncated disk. In the light curve two peaks occur depending on the inclination angle. At large $v_{\rm exp}$, a shallow ``nose'' feature may develop ahead of the loop; its duration depends on the expansion velocity and the inclination angle. We explore the entire interval of black hole spin parameter ranging from extreme prograde to extreme retrograde rotation, $-1<a<1$. In the prograde case, a low-energy tail appears to be more pronounced in the evolving centroid energy of the line. Our results demonstrate the importance to search for X-ray spectral lines in the early phase of TDE flares in order to constrain black hole mass and spin, as well as properties of the innermost accretion flow.

\end{abstract}


\section{Introduction}
\label{sec:ironline}
X-ray spectroscopy reveals that atoms and ions of iron can produce a strong 
Fe K spectral line, which is thought to be shaped by relativistic effects as it emerges 
from the inner regions of an accretion disk.
Hard X-ray photons from a hot black hole corona can irradiate the accretion flow. 
Emission from an irradiated flow then contains the reflection component
\citep[see][and further references cited therein for a review]{fabian_x-ray_2010}. 
The emerging signal includes both the continuum flux and spectral lines, namely,
the prominent fluorescent emission. The
latter is produced by de-excitation of an atom with a vacancy in the
K-shell being filled by an outer electron. The yield of the
fluorescent line scales as $Z^4$ (quartic power of the atomic number),
hence the iron fluorescent line is particularly strong. About 80\% of the 
Fe K line photons come directly from the Fe atoms/ions without experiencing
scattering.

X-ray spectra have been successfully used to study black holes that are surrounded
by an accretion disk. Starving black holes are more difficult to explore. Also when
the surrounding material is inactive (e.g.\ a non-accreting clumpy torus) then the 
electromagnetic signatures cannot be employed to reveal the black hole. In such circumstances 
the immediate vicinity of the black hole horizon can be probed
when a star passes too close to the black hole, so that it becomes destroyed in a Tidal Disruption 
Event (TDE; see \citeauthor{komossa_discovery_1999} \citeyear{komossa_discovery_1999}; 
\citeauthor{gezari_luminous_2009} \citeyear{gezari_luminous_2009}; \citeauthor{kocsis2014} \citeyear{kocsis2014};
\citeauthor{komossa2015} \citeyear{komossa2015},
and further references cited therein). 
The resulting transient accretion is delayed and it occurs much later after the moment of disruption of the star, as the debris material
proceeds to the black hole on the viscous time-scale. Eventually, the gas approaches the innermost stable circular orbit and at this
stage the enhanced accretion gives us an opportunity to examine the black hole by lighting up its environment and producing distinct 
spectral signatures.

While predictions for optical spectra have been discussed in more
detail \citep{strubbe_optical_2009,strubbe_optical_2011,komossa2012}, X-rays reach closer to 
the black hole and relativistically skewed spectral line can strongly constrain the parameters,
thus enabling us to explore effects of strong gravity \citep{cheng2015}.

As the relativistic line is produced near the horizon, the line energy will
be influenced by the gravitational redshift and the Doppler effect
\citep{fabian_x-ray_1989,laor_line_1991}. Hence, the observed spectral profile will
be different from the rest-frame shape of the spectral feature, in particular, the
line centroid energy will be shifted (the change towards higher or lower energy
is possible depending on parameters; 
\citeauthor{karas_theoretical_2006} \citeyear{karas_theoretical_2006}). 
As a result, the reflected spectra
are distorted and the observed iron K$\alpha$ line profile becomes
broad and skewed. Assuming that the flow extends down to the
innermost stable circular orbit \citep[ISCO;][]{bardeen_rotating_1972}, the
emerging spectrum can be used to infer the spin of the central
compact object. This has been done for both Galactic black hole 
X-ray binaries \citep[XRBs; e.g.,][]{miller_nustar_2013} and supermassive
black holes (SMBHs) in Active Galactic Nuclei \citep[AGNs;
e.g.,][]{risaliti_rapidly_2013}. Besides that, the iron line profile can
help us probe the geometry of the hard X-ray emitter
\citep{fabian_long_2002}.

The emergent spectral line depends on the ionization
state of the gaseous material.
As first recognized by \citet{ross_effects_1993}, in the case of X-ray
irradiation the strength of the resulting line emission is
suppressed due to Auger destruction when the ionization parameter
$\xi$ (defined as $\xi \equiv L/nr^2$, where $L$ is the hard X-ray
luminosity, $n$ is the electron number density of the irradiated
disk, and $r$ is the radius) is in the range of $100$--$500$. Then 
the equivalent width of the line
increases for $\xi\gtrsim500$ due to the presence of \ion{Fe}{24}
and \ion{Fe}{25} (the emission lines of iron are dominated by
contributions from these two species). At very high ionization state
($\xi \gtrsim 5000$), the emission line would be suppressed again since a
large fraction of the iron atoms are fully ionized
\citep{matt_iron_1993,matt_iron_1996,karas_cloud_2000}.

A detailed shape of the observed (relativistically skewed) line 
can in principle be used to constrain parameters of the accreting 
system, in particular, the black hole spin. However,
\citet{reynolds_iron_1997} showed that there is a degeneracy in
determining the spin by time-averaged X-ray spectra.
\citet{dovciak_extended_2004} demonstrated that the interplay 
between radius of the inner edge (offset from ISCO) and spin can 
be disentangled, but only with spectra of very high quality.

On the other hand, in a time-resolved process, the
variability of the central hard X-ray sources would ``echo'' on the
irradiated accretion flow. \citet{reynolds_x-ray_1999} and
\citet{young_iron_2000} studied the resulting reverberation features of black
hole spin, given the geometry of the X-ray emitting source, and they
simulated the resulting response to a $\delta$ function flare in AGN. Recently, the promising
reverberation technique has become debated in the context of
time lags emerging in the new X-ray data
\citep{zoghbi2010,zoghbi_discovery_2013}. 

For AGN, the lamp-post model of the irradiated inner accretion disk has been 
adapted as a suitable working hypothesis
\citep{matt91,martocchia_effects_2000,miniutti_light_2004}. In this scenario the
accretion disk is illuminated by a primary source which is located
above the disk plane. The frequently-used assumption places the primary
source on the common axis of the rotating black hole and the
equatorial accretion disk. The
source of primary photons represents a corona or a base of the jet. 
Following an onset of the illumination, the primary photons travel
along null geodesics until some of them hit the surface of the
accretion disk, where they are reprocessed and the emerging
signal (including the reflection spectral line photons) then proceeds
toward a distant observer. In such a geometrical set-up, the
illuminating light front can effectively move at a superluminal speed across the
disk surface. On the other hand, formation of the irradiation source in the 
lamp-post model could be enhanced by relic inactive (collision-less) plasma 
structures persisting as remnants of previous accretion episodes 
\citep{cremaschini2013,cremaschini2014,kovar2014}, including the potential disruption 
of asteroids and small bodies by the supermassive black holes \citep{kostic2009,kostic2012,zubovas2012}.

In the present paper, we investigate a possibility of ionizing the 
freshly accreted material of debris flow, formed from a disrupted or 
partially damaged star as well as pre-existent gas near SMBH. 
This material is activated by X-ray illumination that would be produced 
near the black hole from a 
hot corona or a hot inner accretion flow as the accretion rate grows above 
the Eddington limit.

An expanding 
ionization region develops and proceeds through the accretion flow. 
This is thought to be triggered by a TDE in 
a normal galaxy center, where the induced variation of the spectrum 
can be revealed more easily than in a typical AGN (the latter would be 
intrinsically highly variable). In consequence of the changing ionization 
the spectral line becomes modulated in strength and energy via 
increasing ionization. Indeed, it occurs that the medium near the central
black hole is often in the form of a highly inhomogeneous and clumpy cloudlets which
stay undetected unless illuminated by an external source. 
The actual flare is expected to occur somewhat delayed 
with respect to the pericenter passage of the affected star, as the 
illuminating corona builds gradually during the (super)-Eddington
phase of fast accretion following the moment of TDE.

A non-negligible radial infall velocity of the accreted material is
likely. The ionization front therefore moves radially outward at 
$v=v_{\rm exp}(r)$ not exceeding the speed of light. We set $v_{\rm exp}\simeq {\rm const}$
as one of the free model parameters that could be constrained by the
observed spectra. The scenario of ionization wave being induced by TDE and launched 
from the inner disk provides an alternative scheme with a small number of free 
parameters and somewhat different properties with respect to changing spectral 
properties of the evolving line.

\section{Prospects for Iron Line Reverberation from Tidal Disruption Events}

\subsection{TDEs near SMBHs}
Tidal disruption events (TDEs) can be used to probe the cosmological SMBHs in the centers of inactive
galaxies. A TDE occurs if the stellar orbit falls within radius $r=R_{\rm p}$
such that the ratio between stellar surface gravity and tidal
acceleration at pericenter,
\begin{equation}
  \eta = \left( \frac{R_{\rm p}^3}{GM_\bullet R_\star}
  \frac{GM_\star}{R^2_\star}\right)^{1/2}\lesssim1,
\end{equation}
where $R_{\rm p}$, $R_\star$, $M_\star$ and $M_\bullet$
are the pericenter radius, stellar radius, stellar mass and the central
SMBH mass, respectively
\citep{luminet_tidal_1985,rees_tidal_1988,evans_tidal_1989,phinney_manifestations_1989}. 
The exact value of $\eta$ for the disruption to happen depends on the internal structure,
compressibility and rotation of the star \citep{Eggleton83}. Simulations confirm that, after the disruption, nearly half of the
stellar debris should remain on the bound orbit and they would be
eventually accreted by the SMBH. Based on the viscous mechanism, 
the mass fall-back rate is estimated to be
\begin{equation}
  \dot{M}_{\rm fb}\approx \frac{1}{3}
  \frac{M_{\star}}{t_{\rm fb}}
  \left(\frac{t}{t_{\rm fb}}\right)^{-5/3},
\end{equation}
which can significantly exceed the Eddington accretion rate for a period of
weeks to years \citep{strubbe_optical_2009,strubbe_optical_2011}. The 
mass of the SMBHs associated with TDEs is peaked at $\sim 10^{6.5}$ solar mases 
\citep{stone2014}. The presence of a stellar ring or a secondary intermediate-mass
black hole near 
a SMBH can enhance the TDE rate by increasing the eccentricity fluctuations of stellar
trajectories \citep{vokrouhlicky1998,karas_enhancing_2012,karas2014,li2015}.

In the X-ray and Optical/UV bands, $\sim20$ candidate TDEs have been
reported
\citep{renzini_ultraviolet_1995,bade_detection_1996,komossa_discovery_1999,
  greiner_rx_2000,donley_large-amplitude_2002,
  gezari_ultraviolet_2006,gezari_uv/optical_2008,gezari_luminous_2009,
gezari_ultraviolet-optical_2012,maksym_tidal_2010}. These events are
characterized by  thermal emission with temperature of $\sim
10^4$--$10^5$~K, and the peak bolometric luminosity of about
$10^{43}$--$10^{45}~{\rm erg~s^{-1}}$. For the events with good
coverage during the decay, the luminosity decline was consistent with a
power-law delay with index of $-5/3$. Compared
with quiescent supermassive black holes (SMBHs), such as Sgr A* in
our Galaxy radiating at the level of $10^{35}{~\rm erg~s^{-1}}$, the
luminosity can increase significantly, by several (even up to ten) orders 
of magnitude during TDE outbursts.

In 2011, two putative TDEs were detected: Swift J1644+57
\citep[e.g.,][]{burrows_relativistic_2011,levan_extremely_2011,zauderer_birth_2011}
and Swift J2058.4+0516 \citep[e.g.,][]{cenko_swift_2012}. Their X-ray
spectra showed a power-law shape, and the peak luminosity of
$\sim10^{48}~\rm erg~s^{-1}$ far exceeded the Eddington limit of a
SMBH of a presumed reasonable mass. They also exhibited radio emission with isotropic luminosity of
$\sim 10^{42}~\rm erg~s^{-1}$. These are believed to be due to a
relativistic jet viewed at a very small angle. The amplitude was very
large in these two cases. For Swift J1644+57 the {\it ROSAT} $3\sigma$
upper limit was $2.4\times10^{-13}~\rm erg~s^{-1}~cm^{-2}$, and the
peak flux during the outburst was $\sim 5.0\times 10^{-9}~\rm
erg~s^{-1}~cm^{-2}$ \citep{burrows_relativistic_2011}, showing an
amplitude of more then four orders of magnitude. For Swift
J2058.4+0516, the {\it ROSAT} upper limit was $\sim 10^{-13} ~\rm
erg~s^{-1}~cm^{-2}$ while the peak flux during outburst was $\sim
10^{-10} ~\rm erg~s^{-1}~cm^{-2}$, showing fluctuations with
amplitude of three orders of magnitude.

TDEs can be revealed with a better chance in
sources where the central supermassive black hole is not fed by
vigorous accretion \citep{saxton_preface_2012}. In this context, AGN
have been an unfavored type of objects to reveal tidal
disruptions because accretion is an inherently variable
process which can hide signatures of TDEs. The
X-ray light curve of a typical AGN varies on multiple time-scales,
and so a sporadic flare by a TDE can hardly be discovered in
the background of stochastic fluctuations due to the intrinsic
properties of the accretion flow. Relatively quiet low-luminosity
nuclei appear to be more suitable for this task.

Follow-up optical spectroscopy observations
can give additional information on characteristics to TDE host galaxies
\citep[e.g.,][]{komossa_huge_2004,saxton_tidal_2012,clausen_emission_2012, wang_2012}.
Here we propose that the evolving X-ray spectral line profile expected in the
very early phase of TDE accretion flares can provide a unique probe of the
SMBHs in an otherwise inactive galaxy. Well after the tidal disruption happens, 
an accretion disk forms from the circularized flow of debris. The subsequent evolution is crucial for 
understanding the light curves in TDE events \citep{shen2014}.

The accretion of the debris material starts from low accretion rates and it continues all the way to super-Eddington rates during the rising phase of TDEs. In the early rising phase, the accretion process is expected to proceed in the regimes of radiatively inefficient/hot accretion. The resulting signal mimics the one seen in many AGN where the lamp-post model has been applied to model the irradiation of the disk. A newly formed accretion disk will become ionized when the source reaches and exceeds the Eddington luminosity.

\begin{figure}[tbh]
\includegraphics[width=0.5\textwidth]{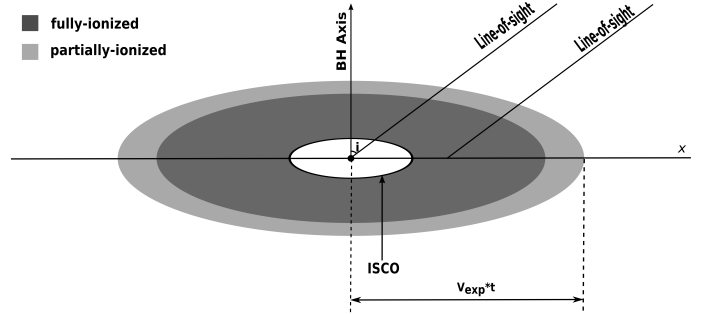}
\caption{A schematic representation of a ``partially ionized
  accretion flow'', where an ionisation wave front is launched from
  around the inner edge and proceeds outwards. The stellar disruption
  causes enhanced accretion onto SMBH which can
  reach and exceed the super-Eddington rate. At time $t$ after 
  the onset of TDE flare, the ionised region expands to radius $r\simeq v_{\rm exp}\,t$. 
  The inner disk is fully ionized and no
  iron line emission can emerge; only the outer part can produce
  the line emission. The fully ionized inner disk and partially
  ionized outer ring are plotted in dark and light gray colors,
  respectively.
  \label{fig:ionf}}
\end{figure}

The idea of this investigation is based on the fact that an instantaneous illumination and
the resulting release of radiation energy by the event occurring near
the center (at the tidal radius $R_{\rm p}$ of the supermassive black hole) shall
induce an ionizing front that propagates outwards and changes the
ionization state of the accretion disk. Consequently, it also
influences the spectral line emissivity in a very specific manner.
The proposed approach offers an interesting
opportunity, nevertheless, it is a challenge because the iron line flux
is probably weak in TDEs, even in underluminous nuclei with a low
accretion rate and no prior accretion disk.

We constructed a simplified scheme to simulate the
response of the fluorescent iron line to the central TDE flare,
where the rising edge of the TDE flare can be the
perfect illuminating signal for reverberation mapping purposes. The
advantage of our description is the parameterization of the model
by only a small number of free parameters. The main purpose of 
this work is to identify characteristic patterns in the line emission, 
not only limited to iron lines, that may be used to probe 
SMBH mass and spin via general relativistic effects in TDEs.
These are thought to be rare events by their nature, however,
the main advantage is that they can be used to explore SMBH in
otherwise non-active galaxies and therefore these systems have
a potential advantage of being more clean than AGN. On the other hand,
different but related aspects of complex iron-line diagnostics have been 
discussed in the context of binary black-hole mergers \citep[e.g.,][]{mckernan2013}.

\subsection{Signatures of Expanding Ionization Front}
\label{sec:ionfront}
In the following we denote the gravitational constant $G$, the
mass of the supermassive black hole $M$, and the speed of light
$c$ (units of radius, time and velocity are $GM/c^2$, $GM/c^3$
and $c$, respectively).

Because in the rising phase of TDE accretion flares the continuum
flux can vary in a large range, we adopt a correspondingly wide range of
ionization into consideration \citep{ross_effects_1993,goosmann2006}. 
For a $10^6~ M_{\odot}$ SMBH, the
peak mass fall-back rate can reach $\sim 1.5 ~ M_{\odot}\,\mbox{yr}^{-1}$
\citep{evans_tidal_1989}, corresponding to the luminosity of
$8.5\times10^{45}~\rm erg~s^{-1}$ with the radiation efficiency of
$0.1$. 

We estimate the
number density of electrons using the radiation
pressure dominated limit of the standard disk
\citep{shakura_black_1973}: $n=1.17\times10^8r^{3/2} ~\rm cm^{-3}$.
The ionization parameter $\xi = L/nr^2=3.3\times10^{15}r^{-7/2} ~\rm
erg~cm~s^{-1}$. Even for $r=1000$, $\xi$ is still as large as
$1.05\times10^5$. As $\xi$ increases inward, the accretion flow
becomes
fully ionized when irradiated by the peak luminosity of the
outburst.

For a particular ring centered at radius $r$ in the equatorial plane,
the ionization parameter at time $t$ is $\xi (t)= L(t-\delta
t)/(nr^2)$, where $L(t')$ denotes the luminosity of the irradiating
emission at time $t'$. The fluorescent emission of the ring is
expected to vary corresponding to the change of the central engine,
but with a delay $\delta t$ which accounts for the photon traveling
from the center to the ring. The reverberating fluorescent emission
emerges at $t=t_0+\delta t$ where $t_0$ is the time of the outburst
onset, and it disappears after $t_1$ when $\xi(t_1) =L(t_1-\delta
t)/(nr^2) \sim 5000$. Looking at the entire disk, during the
outburst rise, the iron line emission is localized to a ring composed
of partially ionized medium. Outside the ring the photons
have not arrived yet; inside the ring the accretion flow is fully
ionized. This partially ionized region expands with time
(Figure~\ref{fig:ionf} illustrates this scheme).

The expansion velocity $v_{\rm exp}$ of the ``partially ionized
ring'' depends on the height $h$ of the illuminating source:
$v_{\rm exp}=\sqrt{1+h^2/r^2}$. If the
source lies near the equatorial plane, $v_{\rm exp}$ is
close to speed of light (for an illuminating source above the equatorial
plane, the expansion would appear even superluminal).
The infall velocity cannot be neglected at high accretion
rates when $v_{\rm exp}$ would be equal $1-v_r$, and hence less than
unity.

\section{Numerical Approach and Simulations}

\subsection{Assumptions and the Model Set-Up}
In the local frame the line-emitting region is confined in a ring area of the disk. In
the simulation we simplified the region to be a ring of narrow width (about unity 
in terms of gravitational radii), and the irradiating source flux evolves 
as a step-function profile. In this scenario the width of the ionization ring relies
on the shape of the rising edge. Then we computed the emerging 
time-resolved iron line profiles, as observed by a distant observer.

For simplicity we assumed a rotating planar disk representing a
geometrically-thin Keplerian accretion disk formed from the TDE
debris material, extending down to ISCO when we see the X-ray emission associated with the TDE. The outer boundary of the
disk in our code was set to be 1000, which is large enough to investigate the current reverberation problem for our purposes. The radial power-law emissivity index was
set to 3 (the Newtonian value at large radii;
\citeauthor{vaughan_xmm-newton_2004} \citeyear{vaughan_xmm-newton_2004}). 
We assumed the iron line emission to
be locally isotropic. Detailed analysis of the angular distribution
of the line emission
\citep{svoboda_role_2009,garcia_improved_2014} showed that it 
is likely to exhibit a weakly limb-brightening dependence, with
maximum ratio of only $\sim$2 compared to an isotropic assumption.
This is less extreme then $I\propto\ln(1+1/\mu)$, which is another
commonly assumed law for the angular dependence of an X-ray
illuminated slab \citep{haardt93}. Hence, our assumption of isotropic emission
would not change the result very much. We assumed the line to be
intrinsically narrow and its profile to be a $\delta$-function in
energy centered at 6.4 keV, i.e., the laboratory energy of
\ion{Fe}{1} K$\alpha$. The results for spectral lines at different
energies can be inferred from our simulation.

With all the assumptions above, the local emissivity is
\begin{eqnarray}
 f(r) = \left\{\begin{array}{ll}  
	{\rm const}*r^{-3},  & v_{\rm exp}*t-1 < r \leq v_{\rm exp}*t  \ \;
	\\[2mm]
	0, \;	     & {\rm otherwise.}	\end{array}\right.\;
\end{eqnarray}
This is obviously a rather simplistic parameterisation of a much more complicated
situation, nonetheless, we find it useful for the sake of demonstrating the
role of GR effects in the context of wave perturbation that is launched
from the inner rim of the disk and then starts travelling outwards.

\subsection{Ray tracing of the emission from an expanding ionization region near a black hole}
The basic
properties of black hole X-ray spectra from the innermost regions of
an accretion disc are well-known
\citep{fabian_broad_2000,reynolds_fluorescent_2003}.
Gravitational effects act on the spectral features from black-hole
accretion disks and coronae by smearing their spectral features and
moving them across energy bins. In this way gravity exerts the
influence on the ultimate form of the observed spectrum
\citep{cunningham_optical_1973,cunningham_effects_1975,karas_theoretical_2006}.
The reprocessed radiation reaches the observer from different regions
of the system. Furthermore, as strong-gravity plays a crucial role,
photons emitted at the same moment may even follow multiple separate paths, joining each other
at the observer at different moments. Individual rays experience
diverse time lags for purely geometrical reasons, for relativistic
time-dilation, and by repressing within the medium.

To simulate the evolution of the observed spectral line,
we have employed the general-relativistic ray-tracing
method (KYcode; \citeauthor{dovciak_xspec_2004}
\citeyear{dovciak_xspec_2004}, \citeyear{dovciak_extended_2004}). This code uses pre-calculated information about the form of light rays (i.e., the geometrical shape and the timing relations along null
geodesics, which describe the photon propagation in the approximation
of geometrical optics). A fine grid is used with
multiple scales that allow us to reproduce the effects of varying
energy shifts, gravitational lensing near caustics, frame-dragging
near the ergosphere, and time delays in the gravitational field of a
rotating (Kerr) black hole (dimensionless spin parameter $-1\leq a
\leq 1$).

\begin{figure*}[tbh]
\centering
\includegraphics[width=0.7\textwidth]{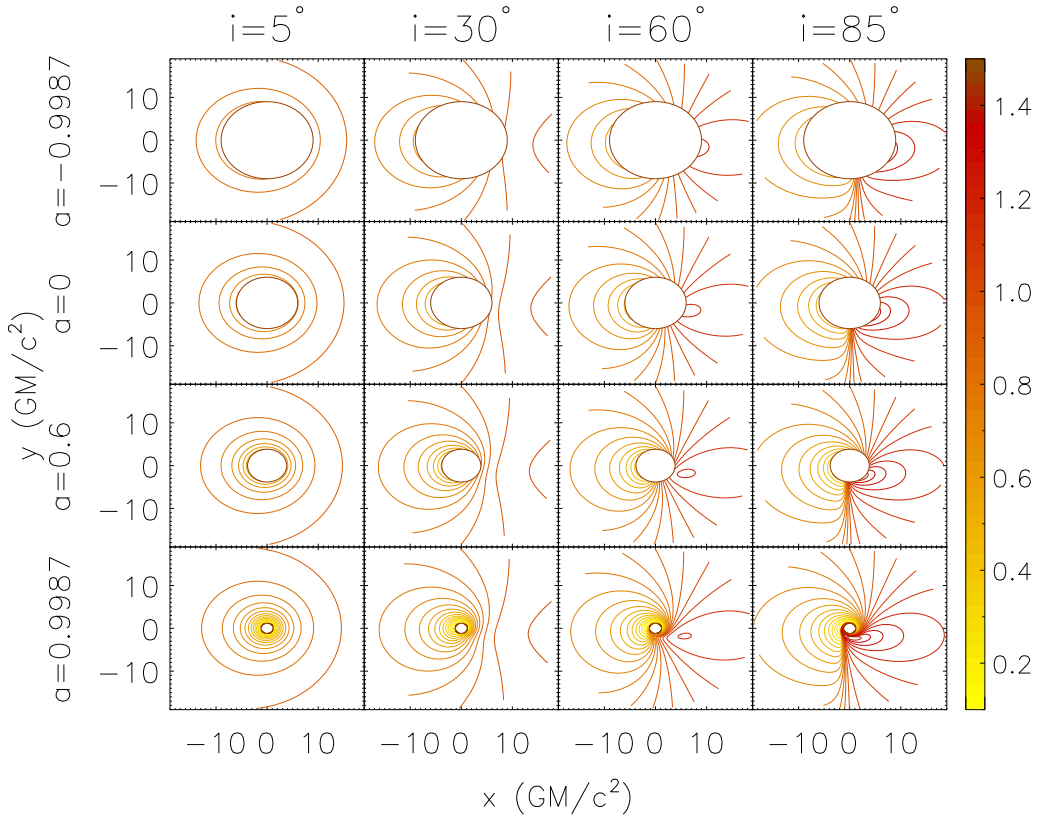} 
\caption{Contour lines of redshift factor $g(r,\phi) \equiv E_{\rm obs}/E_{\rm loc}$ in
the equatorial plane of a rotating black hole for several values of its spin (rows) 
and of observer's inclination (columns). In this plot the black hole rotates in
both clockwise ($a>0$) and counter-clockwise ($a<0$) sense, as indicated
($a=0$ is the static case with zero angular momentum). 
The radiating atoms follow Keplerian
circular motion around the black hole, while a distant observer has
the line-of-sight directed downward. In each panel the black circle in the middle
indicates the location of ISCO. The value of $g$ is color coded, as
indicated by the color bar to the right of the figure.
\label{fig:gfactor}}
\end{figure*}

\begin{figure*}[tbh]
\centering
\includegraphics[width=0.7\textwidth]{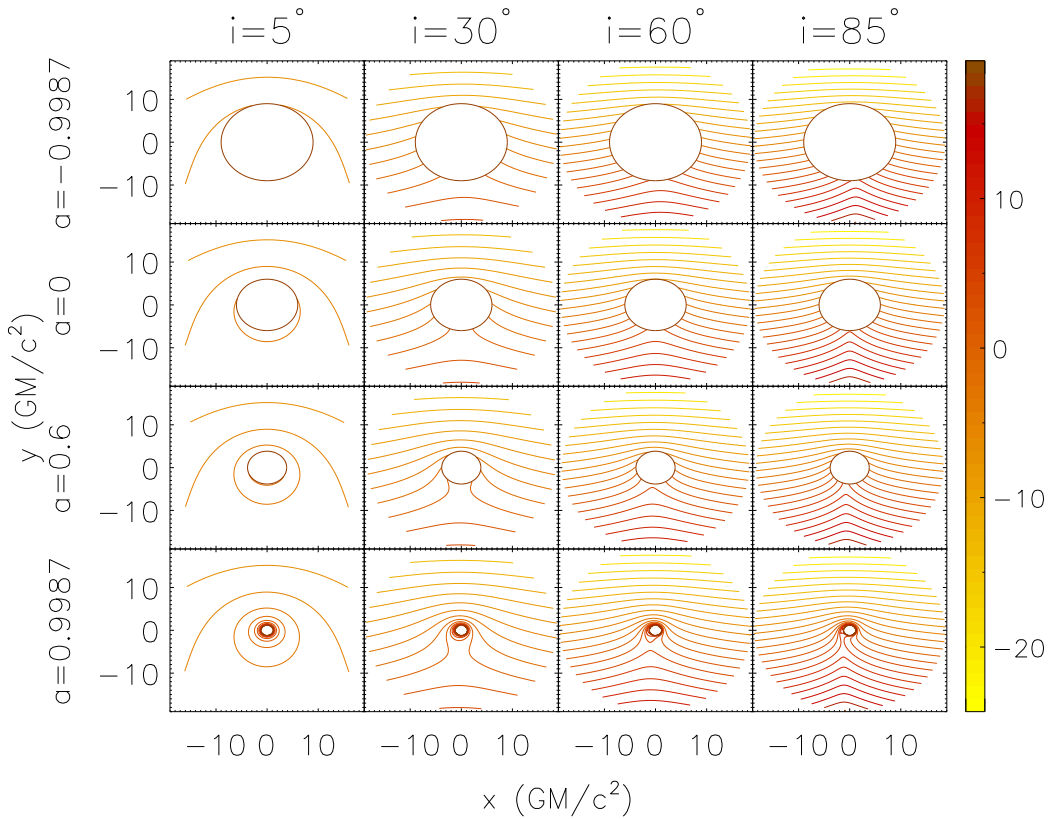}
\caption{Distribution of the relative time delay of photons originating from 
the black hole equatorial plane. The set-up of the plot is the same as in 
Fig.~\ref{fig:gfactor} and the values of the time delay are encoded by the 
color bar (values are given in units of $GM/c^3$).
Notice the prevailing influence of strong gravity near the horizon, where
the light travel time quickly grows above the corresponding Newtonian (flat
space-time) value. This effect distorts the perfectly straight (horizontal)
of the Newtonian time-delay contours which in the relativistic case 
encircle the horizon.  
\label{fig:delay}}
\end{figure*}

First, the source of the emerging signal is 
represented by an expanding ionization wave launched in the equatorial plane,
as introduced in the previous section and further detailed below in
Sec.~\ref{sec:timeresolved}. The resulting spectra and the
corresponding light curves are produced in the form predicted for a
distant observer, i.e., as they are expected to be recorded at the
detector plane (defined by the view-angle inclination with respect to
the black hole rotation axis) at radial infinity. The code uses the
relevant functions stored in FITS files, so that the subsequent computations
of the spectral changes are performed in very efficient manner. This
allows us to reconstruct the impact of energy shifts (both the
Doppler red/blue shifts due to positive/negative component of motion
along the line of sight, as
well as the overall gravitational red shift near the horizon), light
focusing, and the photon travel time from the point of emission
towards an observer far from the source.

We have set-up a grid of the interpolation mesh with respect to
relevant variables: black-hole spin, the inclination angle, and the
expansion velocity of the front. For the spin we select 
exemplary values of $a=-0.998, -0.5, 0$, $0.25$, $0.5$, $0.75$, and $0.998$. 
Both prograde (co-rotating, $a>0$) and retrograde (counter-rotating,
$a<0$) cases are investigated in this work. This covers the entire range of 
spin values that allow for the black hole horizon, while assuming that the 
inner rim is linked with the ISCO radius (the latter recedes in the retrograde case,
and so the relativistic effects become less important).\footnote{In principle, the 
values of $|a|>1$ should not be excluded from the analysis {\em a priori}, 
however, we do not consider this possibility because then a naked singularity develops 
and this would bring us beyond the standard scenario of black-hole accretion (work in 
progress).} For the inclination angle we adopt the grid values in the
range from $i=0$ (corresponding to the rotation axis) up to
$i=85^\circ$ (almost edge-on view), with the resolution of $5^\circ$.
For the expansion velocity, we set the reference value at $v_{\rm
exp}=1$ (corresponding to the case of an illuminating source lying in
the disk plane and the ionization taking effect immediately),
and we show $v_{\rm exp}=0.5$ and 
$0.75$. These allow us to take the effects of accreted debris radial
inflow into account. For reference and comparison we also calculated the case
of $v_{\rm exp}=1.2$ and $2.0$ which can
represent the superluminal expansion for an illuminating source
located above the disc plane, such as the case of lamp-post
scenario in Kerr metric \citep{martocchia_effects_2000}.

Let us note that the constant value of the expansion velocity $v_{\rm exp}={\rm const}$
is introduced to illustrate the expected dependencies and it serves as a 
simplified parameterisation. In a realistic description the ionisation front will
proceed in the interaction with the inflowing material, which exhibits a radially
dependent infall velocity. Also, the relativistic effects will modify the simplistic
scheme. Nevertheless, even the lamp-post scenario finds its phenomenological
substantiation as it can mimic the role of the inner accretion disk self-irradiation,
where the light-bending causes an additional energy input impinging onto the
disk surface predominantly over the location on the disk axis at distance 
of roughly the light circle ($3GM/c^2$ for $a=0$). For this reason we think it is useful
to explore the entire range of $v_{\rm exp}$ from small values to greater than unity.

\section{Results} 
We show how the different values of black hole spin,
inclination angle and expansion velocity affect the evolution of the
relativistic spectral line emission as predicted in this model.

To better understand where the features in the time-resolved line profile
originate, we first illustrate the distribution of redshift
and time delays for photons emitted in the equatorial disk in Kerr
space-time for different values of black-hole spin $a$ and observer's
inclination angle $i$. These functions then determine the time-dependent
profiles of the iron line.

Figure~\ref{fig:gfactor} shows the distribution of the redshift on
the equatorial disk plane, for different values of $a$ and $i$. The redshift 
$g(r,\phi)=E_{\rm obs}/E_{\rm loc}$ is
affected by both gravitational field of the central black hole and
also the Keplerian circular motion of the disk.
The redshift function reaches minimum in the innermost region of
the accretion flow, which is closest to the black hole (see
the bottom panels of Fig.~\ref{fig:gfactor}, where $a=0.998$ and
the ISCO lies on $r=1.23$). The redshift factor can be as small as $\sim0.2$ in the inner region around ISCO.
Photons emitted from this area contribute to the extended red
wing of the broad iron line in X-ray spectra of AGNs and galactic
XRBs. The Doppler effect is more profound for large inclination
angle edge-on systems, as the component of the velocity along the
line-of-sight increases with the inclination. 

Fig.~\ref{fig:delay} shows the effect of purely geometrical (general
relativistic) relative time delay of photons arriving from the
equatorial plane. In the case of
prograde rapidly rotating black hole ($a=0.998$, the bottom panels), the
spectral-line photons experience large time-delays in the inner 
disk. For the high inclination systems ($i=85^\circ$;
right panels), the photons from behind the black hole are delayed
when they pass the black hole. In the extreme case of both large
inclination and rapidly rotating black hole (the right-bottom panel),
the frame-dragging effect is also seen.

\begin{figure*}[tbh] 
\centering
\includegraphics[width=0.9\textwidth]{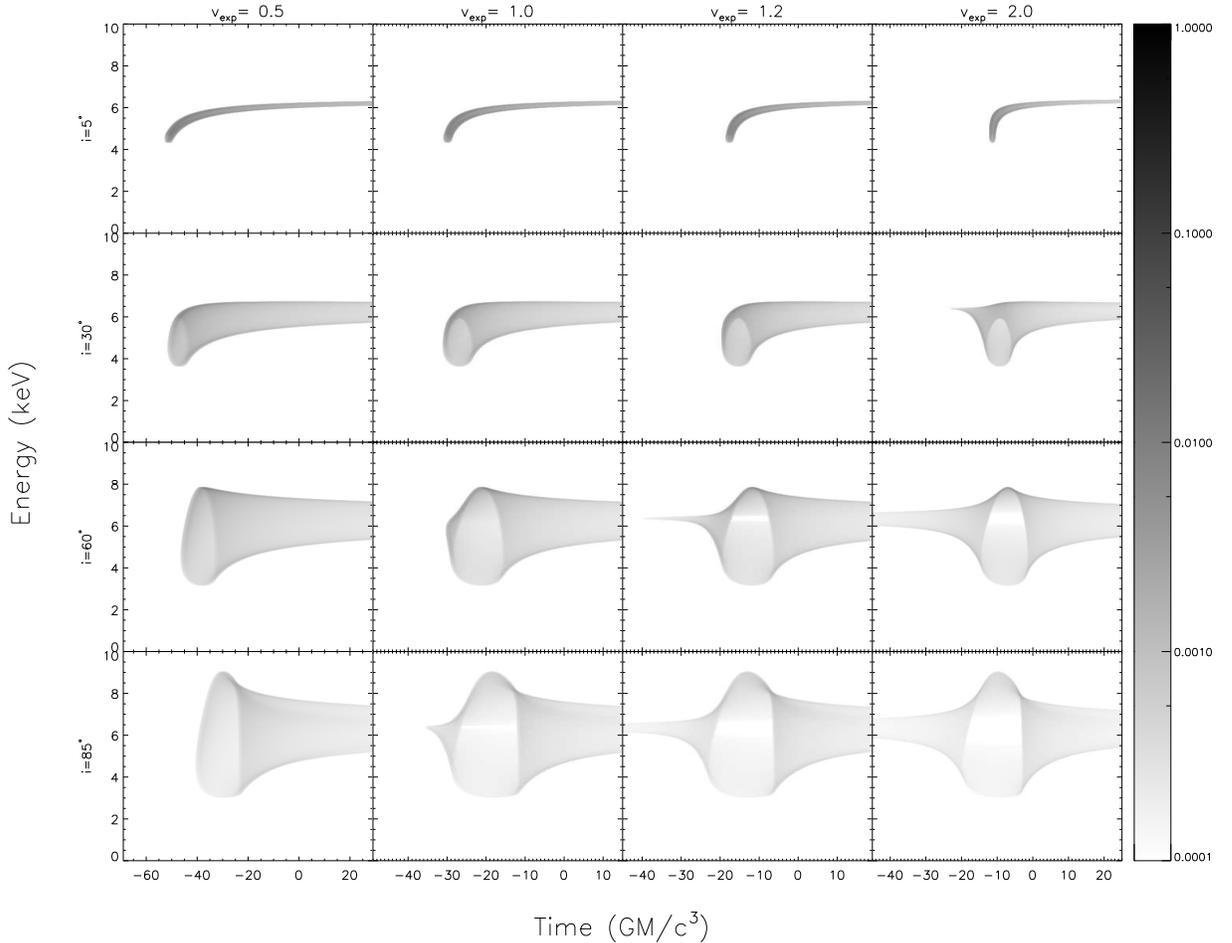}
\caption{Evolution of the observed (computed) iron line photon counts 
(background subtracted) for Schwarzschild black hole ($a=0$), for
different inclination angles and expansion velocities. 
The photon count values are normalized to the maximum value for all panels
and encoded by gray-scale (in arbitrary units). The
corresponding inclination angle and expansion velocity are shown
at left and top axes, respectively.
\label{fig:schwarz}}
\end{figure*}

\begin{figure*}[tbh] 
\centering
\includegraphics[width=0.9\textwidth]{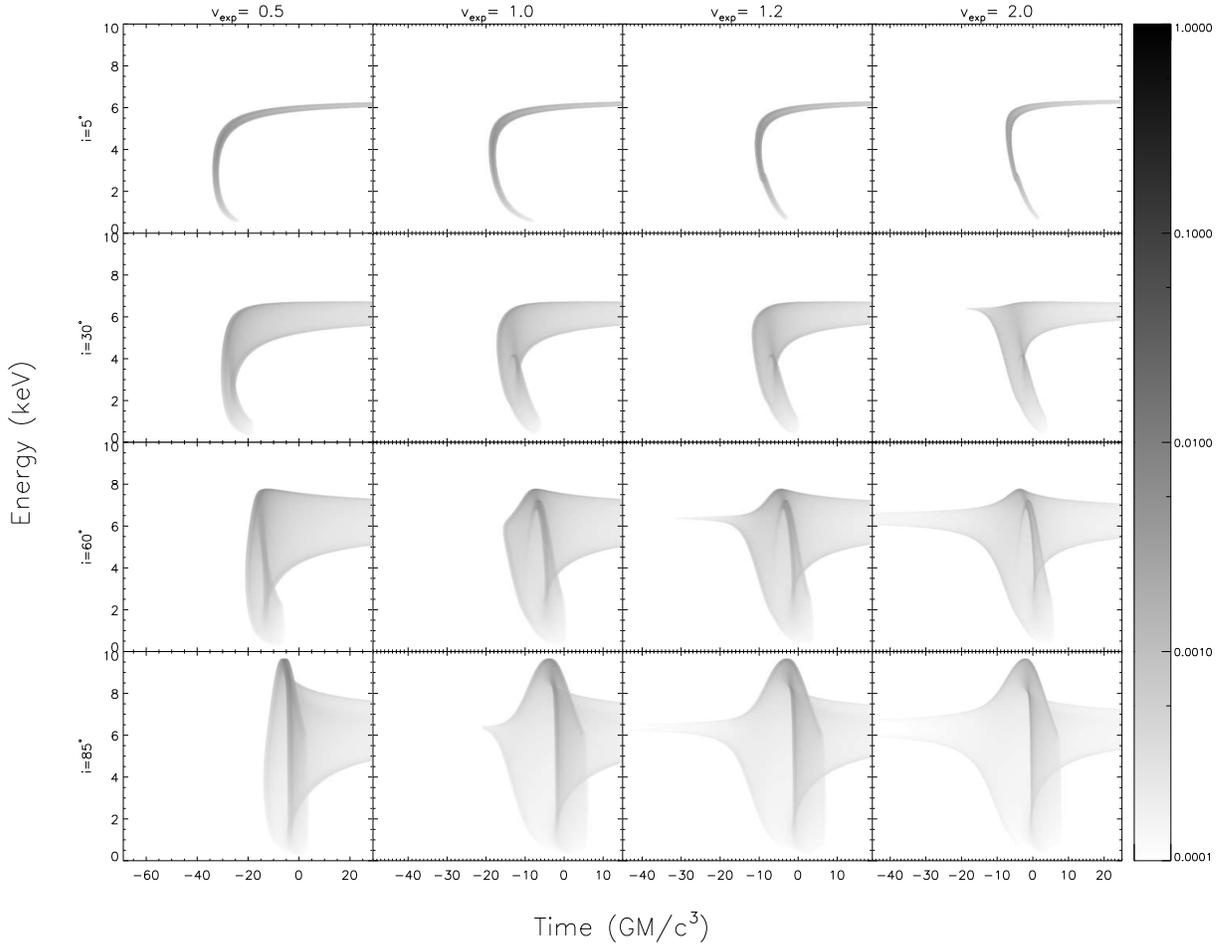}
\caption{As in the previous figure, but for a fast-rotating black hole (spin $a=0.998$). 
Prograde sense of rotation is assumed, i.e., the accreted material co-rotates with the 
black hole.
\label{fig:kerr}}
\end{figure*}

\begin{figure*}[tbh] 
\centering
\includegraphics[width=0.9\textwidth]{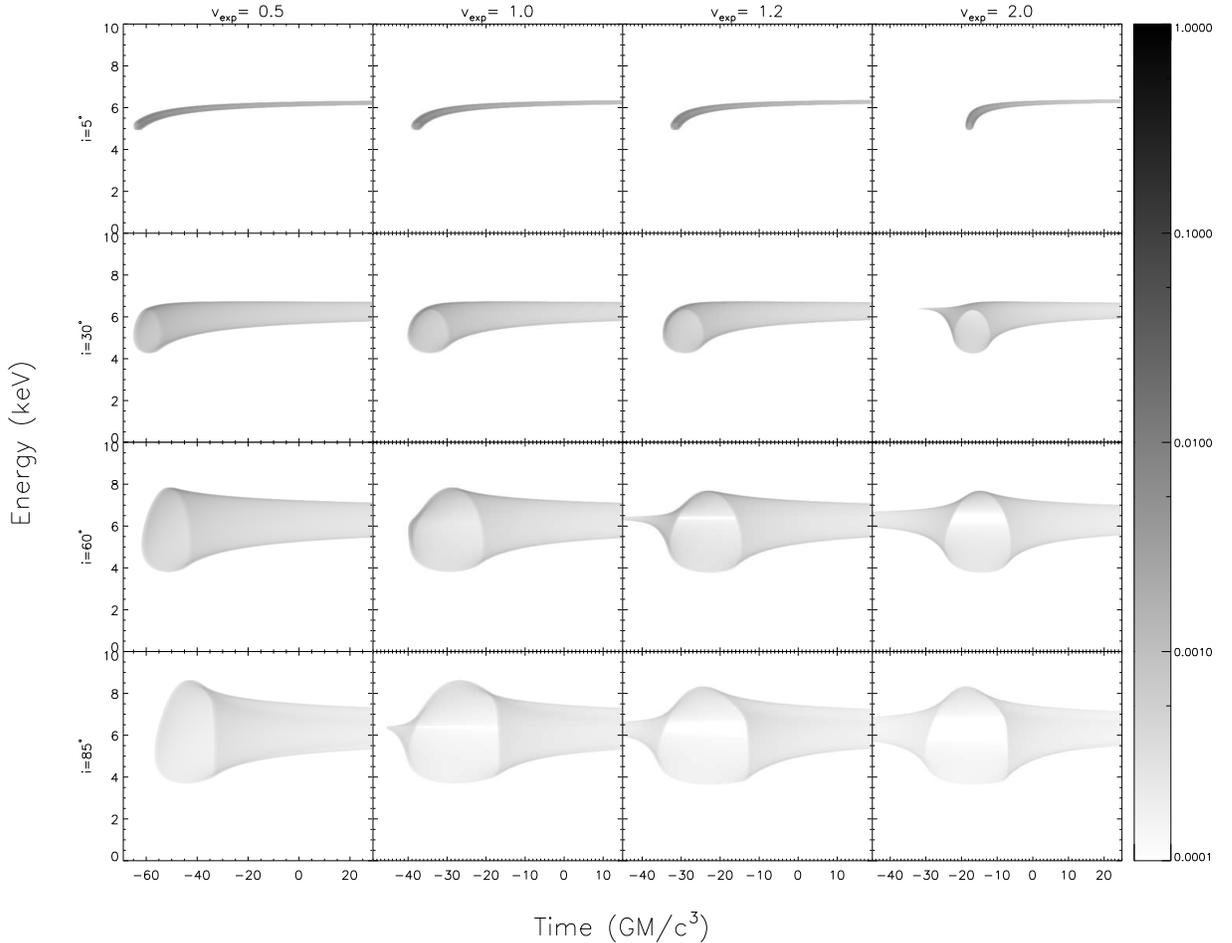}
\caption{As in the previous figure, but for a counter-rotating black hole (spin $a=-0.998$). 
Unlike the previous figure, retrograde rotation means that the accreted material has
its angular momentum opposite to the black hole. This leads to receding ISCO radius, and
therefore decreasing the span of energy that the event covers in the plot above in comparison
with the case of prograde rotation.
\label{fig:kerr_m1}}
\end{figure*}

\begin{figure*}[tbh] 
\centering
\includegraphics[width=0.8\textwidth]{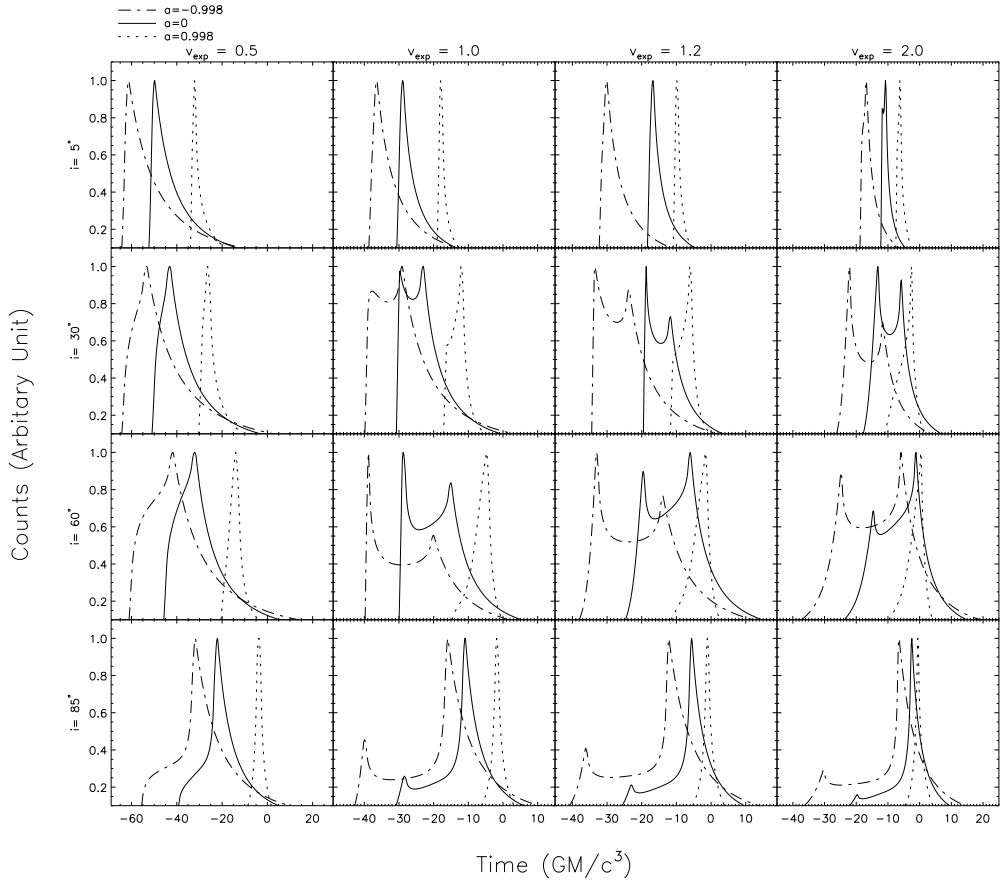}
\caption{The observed count rates in the fluorescent iron line (as computed by our ray-tracing 
  method, background subtracted), for different inclination angles $i$ and
  expansion velocity $v_{\rm exp}$, as indicated at each corresponding row and
  column, respectively. The dash-dotted, solid, and dotted lines correspond
  to the extreme retrograde Kerr ($a=-0.998$), non-rotating Schwarzschild ($a=0$), and extreme prograde Kerr ($a=0.998$) black
  holes, respectively. Each lightcurve is normalized by its maximum value.
  \label{fig:lc}}
\end{figure*}

\label{sec:timeresolved}
In Figures~\ref{fig:schwarz}--\ref{fig:kerr_m1} we show the
time-resolved line profile, both for Schwarzschild ($a=0$) and
extreme Kerr black holes ($a=0.998,-0.998$). We set zero time as mean
time, defined by $t_0 = \int t\,f(t)dt/\int f(t)\,dt$ where $f(t)$ denotes
the photon counts recorded at the detector plane.\footnote{To compare with
observations, it would be natural to define time zero to be the
moment when the onset of the TDE flare reaches the observer.
However in our assumption the
illuminating source lies in the center of the disk plane which is
inside the event horizon, so it is actually not feasible to define
zero time in this way.} The corresponding 
lightcurves are shown in figure~\ref{fig:lc}, where Schwarzschild
and Kerr cases are plotted in dash-dotted and solid lines, respectively. 
Let us note that here we show the (computed) light curves of the 
background-subtracted model; hence, the plot exhibits the 
expected variation of that part of the observed signal restricted to the
energy of the iron line emission.

\begin{figure*}[tbh] 
\centering
\includegraphics[width=0.95\textwidth]{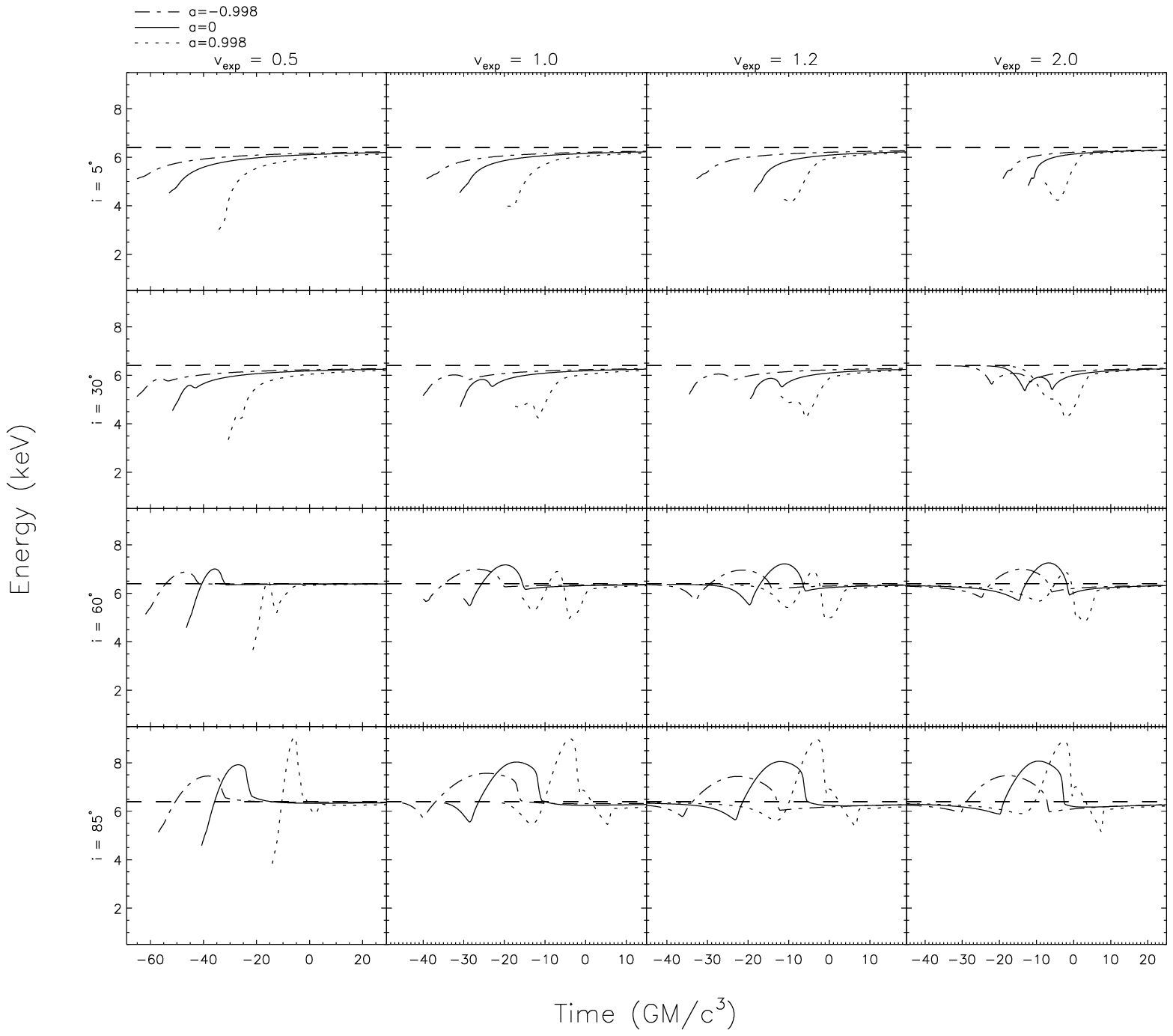}
\caption{The expected time variations of iron line centroid energy $E_{\rm cen}$ for different inclination angles $i$ (rows) and expansion velocities $v_{\rm exp}$ (columns). The line types as in the previous figure: the dash-dotted line for $a=-0.998$), the solid line for $a=0$, and the dotted line for $a=0.998$ cases, respectively. The corresponding inclination angles $i$ and expansion velocities $v_{\rm exp}$ are also given. The rest-frame energy 6.4 keV of \ion{Fe}{1} K$\alpha$ spectral line is indicated by the horizontal dashed lines for the reference.
 \label{fig:cen}}
\end{figure*}

To characterize the changing profile we employ the
line centroid energy, 
\begin{equation}
  E_{\rm cen}=\frac{\int E\,f(E)\,dE} {\int f(E)\,dE},
\end{equation}
where $E$ is the photon energy, and $f(E)$ is the photon number density flux in the
line, both of which are measured with respect to the observer's frame.
While the details of the evolving spectrum are telling about GR effects
from the theoretical point of view, the centroid energy provides a robust
characteristic that can be compared with observation.

For the Schwarzschild black hole case at $i=5^\circ$ (i.e. viewed
almost along the symmetry axis), the iron line has a narrow profile
at any time and the line energy increases with time asymptotically
towards the rest-frame value. As the inclination angle increases
from $5^\circ$ to $85^\circ$ (close to an edge-on view), the observed line
profile at any time becomes broadened due to an increasing velocity component
of the disk material along the line of sight. Once $v_{\rm exp} \geq 0.5$ and $i \geq
30^\circ$, a loop is seen in the line profile vs. time plot (see
Figs.~\ref{fig:schwarz}--\ref{fig:kerr_m1}). The time span of the loop increases with the
inclination angle. At $v_{\rm exp} \gtrsim 1.0$, two peaks can be
seen in the line lightcurve, as shown in Fig.~\ref{fig:lc},
and the first and the second peaks correspond to the left and the
right edges of the loop, respectively. 

We computed the time span of
the two peaks and we found that their time difference stays roughly constant with
varying expansion velocity. For example, when $i=60^\circ$, the time
separation is 13.8, 13.6, 13.4 for $v_{\rm exp}=1.0$, $1.2$ and
$2.0$, respectively. On the other hand, the time separation and
intensity ratio of the two peaks depend on the inclination. The
mean time separation ($v_{\rm exp}=1.0$, $1.2$ and $2.0$) is 7.1,
13.6 and 17.3 for $i=30^\circ$, $60^\circ$, and $85^\circ$. The
intensity ratio of the first to the second peak decreases with
the inclination angle in most cases. During the first peak, a dip occurs in the
centroid-energy evolution, and between the two peaks the centroid
energy exhibits a bump (see Figure~\ref{fig:cen}).

The time delay between the central TDE X-ray irradiating flare and the
induced iron line is denoted as $\Delta t$ and contributed by two factors:
the first one is the expansion time, $t_{\rm exp}\simeq r/v_{\rm exp}$, and
the second one is the travel time needed from the disk to the
observer, $t_{\rm delay}$ (Fig.~\ref{fig:delay}). At
large radius where the spacetime is flat to a good precision,
$t_{\rm delay}\simeq -r\sin\phi+\mbox{const}$, where $\phi$ is the
azimuthal angle on the disk plane. As $v_{\rm exp}$ is of the order
of unity, the two are comparable, and $\Delta t$ is dominated by the
longer one. Hence in the case of $v_{\rm exp} \geq 1$, $\Delta t$ is
dominated by $t_{\rm delay}$. 

At large inclination angles, the
photons originating from the disk in front of the black hole reach
the observer first, then come those photons from the disk near ISCO, and
the last are those photons from the disk behind the black hole. The loop feature 
in the line profile vs. time plot is seen due to the existence of a
hole (due to our assumption that the disk would be truncated inside
the ISCO). The left edge of the loop corresponds to the innermost
isochronous disk plane unaffected by ISCO in front of the black
hole, and the right edge of the loop corresponds to the innermost
isochronous plane unaffected by ISCO behind the black hole.
Roughly, the time span of the loop equals to $l\sim 2r_{\rm ISCO}
\sin i$ (where $r_{\rm ISCO}$ is the radius of the ISCO), which is
consistent with the time separation between the two observed line
intensity peaks in the light curve.

The relative intensity of the second
peak is enhanced for large inclination angles as it is formed
mainly by the photons emerging from the disk behind the black hole (that get more focused at high inclinations). 
In the case of low expansion velocity, $\Delta t$ is dominated by ${t}_{\rm exp}$, 
and the observed image of the emission region expands outwards from the ISCO. In the cases of high expansion velocity combined with high
inclination angle, there are ``noses'' occurring ahead of the loop. In these cases, the observed image of the emission region first expands outside the ISCO and only later on reaches the region below the ISCO. This expansion creates the nose that is followed by the loop and two intensity peaks are visible in the light curve at the time when the image reaches and subsequently leaves the ISCO.

The duration of the nose increases with inclination angle, and
with the expansion velocity as well. The noses are contributed by
photons emerging from the disk in front of the black hole. Higher 
inclination and larger expansion velocity cause that the first image 
of the emission region comes from further away from the centre 
which prolongs the nose in time-energy graphs. The
centroid energy of the noses is close to the rest frame energy of 6.4
keV (this is also reflected in Figs.~\ref{fig:schwarz}--\ref{fig:kerr_m1}). 

Comparing with the Schwarzschild case, in extreme Kerr black hole
cases, the line energy can extend to both lower and higher energies,
due to stronger gravitational and Doppler energy shifts as the inner edge
shrinks with increasing spin. The loop feature is no longer seen, while the
nose stays unaffected. 

At $v_{\rm exp} \gtrsim 0.5$ speed of light, a low-energy tail
appears in the profile (see Fig.~\ref{fig:kerr}). 
A similar signature has been noticed for Kerr black
holes \citep{young_iron_2000}. As the inner edge is very close to the black hole, the photons
from this region of the disk suffer from large time delays (see
bottom panels of Fig.~\ref{fig:delay}) and strong gravitational
redshift (see bottom panels of Fig.~\ref{fig:gfactor}), and for
this region $\Delta t$ is dominated by $t_{\rm delay}$ at $v_{\rm
exp}\geq 0.5$. These two contributing effects produce the delayed low-energy
tail. As a result, the two intensity peaks do not develop in the light curve,
while a dip remains to be seen in the centroid energy evolution.

\section{Discussion and Conclusions}

We simulated the time-resolved iron-line profiles from TDE flares, as 
they are expected to be seen in a distant observer's frame.
Because the maximum mass fall-back rate can be supper-Eddington in
TDEs, a large range of luminosity is expected in these events. Hence, the energetic
photons from the central region can bring the accretion flow to high
ionization state. The rising edge
of TDE flares can produce and modulate the fluorescent iron line. 
The line emission would disappear in the accretion flow where the
disk material is highly ionized (in terms of $\xi\gtrsim5000$).  

We explored a simple parameterization according to which 
the velocity of a moving ionization front $v_{\rm exp}$ defines a different 
kind of variable compared to the more standard 
height $h$ of the lamp-post scenario. 
Other parameters of the model are the black hole spin $a$ (Kerr metric
is assumed) and the line-of-sight view angle $i$ of the system.
Motivation of the model set-up arises in the context of spectral-line 
reverberation mapping of the spectral line from the debris disk
illuminated after a tidal-disruption event. We suggest that this scheme
can be useful to study the properties of tentative TDEs, where the
emerging spectral line is modified by an outward propagating circle which 
grows and changes the line
emissivity. The accretion process is expected to be highly non-stationary in such a situation. 

We conclude that notable signatures can be expected:

(i)---For $i \gtrsim 30^\circ$, $v_{\rm exp} \gtrsim1$, and small 
  or negative (retrograde) spin, a loop feature 
  appears in iron line profile vs.\ time plot and it manifests itself in
  the line flux showing two peaks in the corresponding light curve.
  The time separation and intensity ratio of the two peaks mainly
  depend on the inclination angle. For large values of 
  positive (prograde) spin parameter, the radiation 
   emitted close above the horizon destroys the loop (see also the point (iii) below).
	
(ii)---For large expansion velocities of the ionization circle and
	large inclination angles, a nose appears ahead of the loop in the
	iron line profile vs. time plot. The duration of the nose is
	determined by
	both the expansion velocity and the inclination angle.
  
(iii)---For $v_{\rm exp} \gtrsim 0.5$ and $a\rightarrow1$ (extreme
	prograde rotation), a more prominent low-energy tail is 
	seen. Occurrence of this feature can serve as a probe 
	of the fast-rotating Kerr black hole. On the other hand, in the 
	non-rotating and retrograde cases the low-energy tail is suppressed 
	and it happens earlier in time.

Let us note that the retrograde sense of rotation is often dismissed in accretion 
disk modelling and spectra fitting because it appears to be less likely compared 
to the prograde case. It can be argued that the long-term evolution the accretion 
flow tends to align the angular momentum vector of the flow with that of the black
hole. However, the situation with TDEs is quite different \citep{sochora2011}. An infalling star approaches
the central SMBH from the surrounding nuclear star-cluster, which can be considered 
as spherical and isotropic in zero approximation, and so counter-rotating motion
is equally possible as co-rotating with the black hole. Furthermore, counter-rotating
stellar rings and disks have been reported in some galaxies, so we included
this possibility in our considerations.

Once a TDE-induced relativistic line is detected and its parameters 
reliably determined in the future,
these signatures in time-resolved signal can be used to constrain the
black hole spin, inclination, and the expansion velocity of the ionization front. 

\acknowledgments 
We would like to thank the anonymous referee for useful comments. This work was supported in part by the National Natural Science Foundation of China under grant No. 11073043, 11333005, and 11350110498, by Strategic Priority Research Programme ``The Emergence of Cosmological Structures'' under grant No. XDB09000000, and the XTP project No. XDA04060604, by the Shanghai Astronomical Observatory Key Project, and by the Chinese Academy of Sciences Fellowship for Young International Scientists Grant. VK thanks to Czech Science Foundation -- Deutsche Forschungsgemeinschaft project (GA\v{C}R 13-00070J) and the scientific exchange programme M\v{S}MT--Kontakt (LH14049), titled ``Spectral and Timing Properties of Cosmic Black Holes''. MD acknowledges continued support from EU 7th Framework Programme No.\ 312789 ``StrongGravity''.


\begin{thebibliography}{}
\expandafter\ifx\csname natexlab\endcsname\relax\def\natexlab#1{#1}\fi

\bibitem[{Bade {et~al.}(1996)Bade, Komossa, \&
  Dahlem}]{bade_detection_1996} Bade, N., Komossa, S., \& Dahlem, M.
  1996, \aap, 309, L35

\bibitem[{Bardeen {et~al.}(1972)Bardeen, Press, \&
  Teukolsky}]{bardeen_rotating_1972} Bardeen, J.~M., Press, W.~H., \&
  Teukolsky, S.~A. 1972, \apj, 178, 347

\bibitem[{Burrows {et~al.}(2011)Burrows, Kennea, Ghisellini, Mangano,
  Zhang, Page, Eracleous, Romano, Sakamoto, Falcone, Osborne,
Campana, Beardmore, Breeveld, Chester, Corbet, Covino, Cummings,
D'Avanzo, D'Elia, Esposito, Evans, Fugazza, Gelbord, Hiroi, Holland,
Huang, Im, Israel, Jeon, Jeon, Jun, Kawai, Kim, Krimm, Marshall,
Mészáros, Negoro, Omodei, Park, Perkins, Sugizaki, Sung, Tagliaferri,
Troja, Ueda, Urata, Usui, Antonelli, Barthelmy, Cusumano, Giommi,
Melandri, Perri, Racusin, Sbarufatti, Siegel, \&
Gehrels}]{burrows_relativistic_2011} Burrows, D.~N., Kennea, J.~A.,
Ghisellini, G., {et~al.} 2011, \nat, 476, 421

\bibitem[{Cenko {et~al.}(2012)Cenko, Krimm, Horesh, Rau, Frail,
  Kennea, Levan, Holland, Butler, Quimby, Bloom, Filippenko, Gal-Yam,
Greiner, Kulkarni, Ofek, Olivares~E, Schady, Silverman, Tanvir, \&
Xu}]{cenko_swift_2012} Cenko, S.~B., Krimm, H.~A., Horesh, A.,
{et~al.} 2012, \apj, 753, 77

\bibitem[{Chen \& Bogdanovi\'c(2015)}]{cheng2015} Cheng, R. M., \& 
 Bogdanovi\'c, T.\ 2015, \prd, 90, id. 064020

\bibitem[Clausen et al.(2012)]{clausen_emission_2012} Clausen, D.,
 Eracleous, M., Sigurdsson, S., \& Irwin, J.~A.\ 2012, EPJ Web of Conf., 39, 1005 

\bibitem[Cremaschini et al.(2013)]{cremaschini2013} Cremaschini, C., 
 Kov{\'a}{\v r}, J., Slan{\'y}, P., Stuchl{\'{\i}}k, Z., \& Karas, V.\ 2013, \apjs, 209, 15

\bibitem[Cremaschini \& Stuchl\'{\i}k (2014)]{cremaschini2014} Cremaschini, C.,
 \& Stuchl\'{\i}k, Z.  2014, Physics of Plasmas, 21, 042902

\bibitem[{Cunningham(1975)}]{cunningham_effects_1975} Cunningham,
  C.~T. 1975, \apj, 202, 788

\bibitem[{Cunningham \& Bardeen(1973)}]{cunningham_optical_1973}
  Cunningham, C.~T., \& Bardeen, J.~M. 1973, \apj, 183, 237

\bibitem[{Donley {et~al.}(2002)Donley, Brandt, Eracleous, \&
  Boller}]{donley_large-amplitude_2002} Donley, J.~L., Brandt, W.~N.,
  Eracleous, M., \& Boller, T. 2002, \aj, 124,
  1308

\bibitem[{Dov{\v c}iak {et~al.}(2004{\natexlab{a}})Dov{\v c}iak, Karas,
  Martocchia, Matt, \& Yaqoob}]{dovciak_xspec_2004} Dov{\v c}iak, M.,
  Karas, V., Martocchia, A., Matt, G., \& Yaqoob, T.
  2004{\natexlab{a}}, in proceedings of RAGtime: Workshops on 
  Black Holes and Neutron Stars, eds. S. Hled\'\i k and Z. Stuchl\'\i k (Opava: Czech Republic), 
  pp. 33--73 (astro-ph/0407330)

\bibitem[{Dov{\v c}iak {et~al.}(2004{\natexlab{b}})Dov{\v c}iak, Karas, \&
  Yaqoob}]{dovciak_extended_2004} Dov{\v c}iak, M., Karas, V., \& Yaqoob,
  T. 2004{\natexlab{b}}, \apjs, 153, 205
  
\bibitem[{Eggleton(1983)Eggleton}]{Eggleton83}Eggleton, P. P. 1983, \apj, 286, 368

\bibitem[{Evans \& Kochanek(1989)}]{evans_tidal_1989} Evans, C.~R.,
  \& Kochanek, C.~S. 1989, \apjl, 346, L13

\bibitem[{Fabian {et~al.}(2000)Fabian, Iwasawa, Reynolds, \&
  Young}]{fabian_broad_2000} Fabian, A.~C., Iwasawa, K., Reynolds,
  C.~S., \& Young, A.~J. 2000, \pasp, 112, 1145

\bibitem[{Fabian {et~al.}(1989)Fabian, Rees, Stella, \&
  White}]{fabian_x-ray_1989} Fabian, A.~C., Rees, M.~J., Stella, L.,
  \& White, N.~E. 1989, \mnras, 238, 729

\bibitem[{Fabian \& Ross(2010)}]{fabian_x-ray_2010} Fabian, A.~C., \&
  Ross, R.~R. 2010, \ssr, 157, 167

\bibitem[{Fabian {et~al.}(2002)Fabian, Vaughan, Nandra, Iwasawa,
  Ballantyne, Lee, De~Rosa, Turner, \& Young}]{fabian_long_2002}
  Fabian, A.~C., Vaughan, S., Nandra, K., {et~al.} 2002, \mnras, 335,
  L1

\bibitem[{Garcia {et~al.}(2014)Garcia, Dauser, Lohfink, Kallman,
  Steiner, McClintock, Brenneman, Wilms, Eikmann, Reynolds, \&
Tombesi}]{garcia_improved_2014} Garcia, J., Dauser, T., Lohfink, A.,
{et~al.} 2014, \apj, 782, 76

\bibitem[{Gezari {et~al.}(2006)Gezari, Martin, Milliard, Basa,
  Halpern, Forster, Friedman, Morrissey, Neff, Schiminovich, Seibert,
Small, \& Wyder}]{gezari_ultraviolet_2006} Gezari, S., Martin, D.~C.,
Milliard, B., {et~al.} 2006, \apj, 653, L25

\bibitem[{Gezari {et~al.}(2008)Gezari, Basa, Martin, Bazin, Forster,
  Milliard, Halpern, Friedman, Morrissey, Neff, Schiminovich,
Seibert, Small, \& Wyder}]{gezari_uv/optical_2008} Gezari, S., Basa,
S., Martin, D.~C., {et~al.} 2008, \apj, 676, 944

\bibitem[{Gezari {et~al.}(2009)Gezari, Heckman, Cenko, Eracleous,
  Forster, Gonçalves, Martin, Morrissey, Neff, Seibert, Schiminovich,
\& Wyder}]{gezari_luminous_2009} Gezari, S., Heckman, T., Cenko,
S.~B., {et~al.} 2009, \apj, 698, 1367

\bibitem[{Gezari {et~al.}(2012)Gezari, Chornock, Rest, Huber,
  Forster, Berger, Challis, Neill, Martin, Heckman, Lawrence, Norman,
Narayan, Foley, Marion, Scolnic, Chomiuk, Soderberg, Smith, Kirshner,
Riess, Smartt, Stubbs, Tonry, Wood-Vasey, Burgett, Chambers, Grav,
Heasley, Kaiser, Kudritzki, Magnier, Morgan, \&
Price}]{gezari_ultraviolet-optical_2012} Gezari, S., Chornock, R.,
Rest, A., {et~al.} 2012, \nat, 485, 217

\bibitem[Goosmann et al.(2006)]{goosmann2006} Goosmann, R. W., Czerny, B., Mouchet, M., Ponti, G., et al.\
 2006, \aap, 454, 741

\bibitem[{Greiner {et~al.}(2000)Greiner, Schwarz, Zharikov, \&
  Orio}]{greiner_rx_2000} Greiner, J., Schwarz, R., Zharikov, S., \&
  Orio, M. 2000, \aap, 362, L25
  
\bibitem[{Haardt (1993)}]{haardt93}Haardt, F. 1993, 413, 680

\bibitem[{Karas {et~al.}(2000)Karas, Czerny, Abrassart, \&
  Abramowicz}]{karas_cloud_2000}
  Karas, V., Czerny, B., Abrassart, A., \& Abramowicz, M.~A.
  2000, \mnras, 318, 547

\bibitem[{Karas(2006)}]{karas_theoretical_2006} Karas, V. 2006,
  AN, 327, 961

\bibitem[Karas {et al.}(2014)]{karas2014} Karas, V., Dov\v{c}iak, M., Kunneriath, D., Yu, W., \& Zhang, W.
 2014, in proceedings of RAGtime: Workshops on Black Holes and Neutron Stars, eds. Z.~Stuchl\'{\i}k, 
 G.~T\"or\"ok, \& T.~Pech\'a\v{c}ek (Opava, Czech Republic), in press (arXiv:1409.3746)

\bibitem[Karas \& {\v S}ubr(2012)]{karas_enhancing_2012} Karas, V., \& {\v S}ubr,
	L.\ 2012, EPJ Web of Conf., 39, 1003 

\bibitem[Kocsis \& Loeb(2014)]{kocsis2014} Kocsis, B., \& Loeb, A.\ 2014,
 Space Science Reviews, 183, 163

\bibitem[Komossa(2012)]{komossa2012} Komossa, S. 2012, in proceedings of Tidal Disruption 
 Events and AGN Outbursts (Madrid, Spain), eds.\ R. Saxton \& S. Komossa, 
 EPJ Web of Conf., vol. 39, id.~02001

\bibitem[Komossa(2015)]{komossa2015} Komossa, S. 2015, in proceedings of 
Swift: 10 Years of Discovery (Rome, Italy), Journal of High-Energy Astrophysics,
in press (arXiv:1505.01093) 

\bibitem[{Komossa \& Greiner(1999)}]{komossa_discovery_1999} Komossa,
  S., \& Greiner, J. 1999, \aap, 349, L45

\bibitem[{Komossa {et~al.}(2004)Komossa, Halpern, Schartel, Hasinger,
  Santos-Lleo, \& Predehl}]{komossa_huge_2004} Komossa, S., Halpern,
  J., Schartel, N., {et~al.} 2004, \apjl, 603, L17
  
\bibitem[Kosti{\'c} et al.(2009)]{kostic2009} Kosti{\'c}, U., {\v C}ade{\v z}, A., Calvani, M., \& Gomboc, A.\ 2009, \aap, 496, 307
  
\bibitem[Kosti{\'c} et al.(2012)]{kostic2012} Kosti{\'c}, U., {\v 
C}ade{\v z}, A., Calvani, M., \& Gomboc, A.\ 2012, EPJ Web of Conf., 39, 07004  
  
\bibitem[Kov{\'a}{\v r} et al.(2014)]{kovar2014} Kov{\'a}{\v r}, 
J., Slan{\'y}, P., Cremaschini, C., et al.\ 2014, \prd, 90, 044029   

\bibitem[{Laor(1991)}]{laor_line_1991} Laor, A. 1991, \apj, 376, 90

\bibitem[Li et al.(2015)]{li2015} Li, G., Naoz, S., Kocsis, B., \& Loeb, A. 2015, \mnras, submitted (arXiv:1502.03825)

\bibitem[{Luminet \& Marck(1985)}]{luminet_tidal_1985}
  Luminet, J.-P., \& Marck, J.-A. 1985, \mnras, 212, 57

\bibitem[{Levan {et~al.}(2011)Levan, Tanvir, Cenko, Perley, Wiersema,
  Bloom, Fruchter, Postigo, O'Brien, Butler, van~der Horst, Leloudas,
  Morgan, Misra, Bower, Farihi, Tunnicliffe, Modjaz, Silverman,
  Hjorth, Thone, Cucchiara, Cerón, Castro-Tirado, Arnold, Bremer,
  Brodie, Carroll, Cooper, Curran, Cutri, Ehle, Forbes, Fynbo,
  Gorosabel, Graham, Hoffman, Guziy, Jakobsson, Kamble, Kerr,
  Kasliwal, Kouveliotou, Kocevski, Law, Nugent, Ofek, Poznanski,
  Quimby, Rol, Romanowsky, Sánchez-Ramírez, Schulze, Singh, van
  Spaandonk, Starling, Strom, Tello, Vaduvescu, Wheatley, Wijers,
  Winters, \& Xu}]{levan_extremely_2011} Levan, A.~J., Tanvir, N.~R.,
  Cenko, S.~B., {et~al.} 2011, Science, 333, 199

\bibitem[{Maksym {et~al.}(2010)Maksym, Ulmer, \&
  Eracleous}]{maksym_tidal_2010} Maksym, W.~P., Ulmer, M.~P., \&
  Eracleous, M. 2010, \apj, 722, 1035

\bibitem[{Martocchia {et~al.}(2000)Martocchia, Karas, \&
  Matt}]{martocchia_effects_2000} Martocchia, A., Karas, V., \& Matt,
  G. 2000, \mnras, 312, 817

\bibitem[{Matt {et~al.}(1993)Matt, Fabian, \& Ross}]{matt_iron_1993}
  Matt, G., Fabian, A.~C., \& Ross, R.~R. 1993, \mnras, 262, 179

\bibitem[Matt et al.(1991)]{matt91} Matt, G., Perola, G.~C., \& Piro, L.\ 1991, \aap, 247, 25

\bibitem[{Matt {et~al.}(1996)Matt, Fabian, \& Ross}]{matt_iron_1996}
  ---. 1996, \mnras, 278, 1111

\bibitem[McKernan et al.(2013)]{mckernan2013}McKernan, B., Ford, K. E. S., Kocsis, B., \& Haiman, Z.  2013, \mnras, 432, 1468

\bibitem[{Miller {et~al.}(2013)Miller, Parker, Fuerst, Bachetti,
  Harrison, Barret, Boggs, Chakrabarty, Christensen, Craig, Fabian,
Grefenstette, Hailey, King, Stern, Tomsick, Walton, \&
Zhang}]{miller_nustar_2013} Miller, J.~M., Parker, M.~L., Fuerst, F.,
{et~al.} 2013, \apjl, 775, L45

\bibitem[{Miniutti \& Fabian(2004)}]{miniutti_light_2004} Miniutti,
  G., \& Fabian, A.~C. 2004, \mnras, 349, 1435

\bibitem[{Phinney(1989)}]{phinney_manifestations_1989} Phinney, E.~S. 1989,
in procedings of IAU Symp., vol. 136, The Center of the Galaxy, ed. {M.~Morris} (Kluwer Academic Publishers, Dordrecht), p.\ 543

\bibitem[{Rees(1988)}]{rees_tidal_1988} Rees, M.~J. 1988, \nat, 333, 523

\bibitem[{Renzini {et~al.}(1995)Renzini, Greggio,
  di~Serego~Alighieri, Cappellari, Burstein, \&
Bertola}]{renzini_ultraviolet_1995} Renzini, A., Greggio, L.,
di~Serego~Alighieri, S., {et~al.} 1995, \nat, 378, 39

\bibitem[{Reynolds \& Begelman(1997)}]{reynolds_iron_1997} Reynolds,
  C.~S., \& Begelman, M.~C. 1997, \apj, 488, 109

\bibitem[{Reynolds \& Nowak(2003)}]{reynolds_fluorescent_2003}
  Reynolds, C.~S., \& Nowak, M.~A. 2003, \physrep, 377, 389

\bibitem[{Reynolds {et~al.}(1999)Reynolds, Young, Begelman, \&
  Fabian}]{reynolds_x-ray_1999} Reynolds, C.~S., Young, A.~J.,
  Begelman, M.~C., \& Fabian, A.~C. 1999, \apj, 514, 164

\bibitem[{Risaliti {et~al.}(2013)Risaliti, Harrison, Madsen, Walton,
  Boggs, Christensen, Craig, Grefenstette, Hailey, Nardini, Stern, \&
Zhang}]{risaliti_rapidly_2013} Risaliti, G., Harrison, F.~A., Madsen,
K.~K., {et~al.} 2013, \nat, 494, 449

\bibitem[{Ross \& Fabian(1993)}]{ross_effects_1993} Ross, R.~R., \&
  Fabian, A.~C. 1993, \mnras, 261, 74


\bibitem[Saxton \& Komossa(2012)]{saxton_preface_2012} Saxton, R., \& Komossa,
 S.\ 2012, EPJ Web of Conf., 39, id.\ 1 

\bibitem[{Saxton {et~al.}(2012)Saxton, Read, Esquej, Komossa,
  Dougherty, Rodriguez-Pascual, \& Barrado}]{saxton_tidal_2012}
  Saxton, R.~D., Read, A.~M., Esquej, P., {et~al.} 2012, \aap, 541, 106

\bibitem[{Shakura \& Sunyaev(1973)}]{shakura_black_1973}
Shakura, N.~I., \& Sunyaev, R.~A. 1973, \aap, 24, 337

\bibitem[Shen \& Matzner(2014)]{shen2014} Shen, R.-F., \& Matzner, C.~D.\ 2014, \apj, 784, 87

\bibitem[Sochora et al.(2011)]{sochora2011} Sochora, V., Karas, V., Svoboda, J., \& Dov\v{c}iak, M.\ 2011, \mnras, 418, 276

\bibitem[Stone \& Metzger(2014)]{stone2014} Stone, N.~C., \& Metzger, B.~D.\ 2014, \mnras, submitted (arXiv:1410.7772)

\bibitem[{Strubbe \& Quataert(2009)}]{strubbe_optical_2009} Strubbe, L.~E., \& Quataert, E. 2009, \mnras, 400, 2070

\bibitem[{Strubbe \& Quataert(2011)}]{strubbe_optical_2011} ---. 2011, \mnras, 415, 168

\bibitem[{Svoboda {et~al.}(2009)Svoboda, Dov{\v c}iak, Goosmann, \&
  Karas}]{svoboda_role_2009} Svoboda, J., Dov{\v c}iak, M., Goosmann, R.,
  \& Karas, V. 2009, \aap, 507, 1

\bibitem[{Vaughan {et~al.}(2004)Vaughan, Fabian, Ballantyne, De~Rosa,
  Piro, \& Matt}]{vaughan_xmm-newton_2004} Vaughan, S., Fabian,
  A.~C., Ballantyne, D.~R., {et~al.} 2004, \mnras, 351, 193
  
\bibitem[Vokrouhlick\'y \& Karas(1998)]{vokrouhlicky1998}Vokrouhlick\'y, D., \& Karas, V.\ 1998, \mnras, 298, 53
  
\bibitem[Wang et al.(2012)]{wang_2012} Wang, T.-G., Zhou, H.-Y., 
Komossa, S., et al.\ 2012, \apj, 749, 115

\bibitem[{Young \& Reynolds(2000)}]{young_iron_2000} Young, A.~J., \&
  Reynolds, C.~S. 2000, \apj, 529, 101

\bibitem[{Zauderer {et~al.}(2011)Zauderer, Berger, Soderberg, Loeb,
  Narayan, Frail, Petitpas, Brunthaler, Chornock, Carpenter, Pooley,
Mooley, Kulkarni, Margutti, Fox, Nakar, Patel, Volgenau, Culverhouse,
Bietenholz, Rupen, Max-Moerbeck, Readhead, Richards, Shepherd, Storm,
\& Hull}]{zauderer_birth_2011} Zauderer, B.~A., Berger, E.,
Soderberg, A.~M., {et~al.} 2011, \nat, 476, 425

\bibitem[{Zoghbi {et~al.}(2010)}]{zoghbi2010} Zoghbi, A., Fabian, A. C., Uttley, P., Miniutti, G., et al.\
2010, \mnras, 401, 2419

\bibitem[{Zoghbi {et~al.}(2013)Zoghbi, Reynolds, Cackett, Miniutti,
  Kara, \& Fabian}]{zoghbi_discovery_2013} Zoghbi, A., Reynolds, C.,
  Cackett, E.~M., {et~al.} 2013, \apj, 767, 121
  
\bibitem[{Zubovas {et al.}(2012)Zubovas, Nayakshin, \& Markoff}]{zubovas2012}
 Zubovas, K., Nayakshin, S., \& Markoff, S. 2012, MNRAS, 421, 1315
 
\end{thebibliography}
\end{document}